\documentclass[prb,twocolumn,showpacs]{revtex4}
\usepackage{graphicx} 
\usepackage{amsmath,amsthm,amssymb,mathrsfs}
\usepackage{bm}

\begin{document}

\title{Phase dynamics of oscillating magnetizations coupled via spin pumping}

\author{Tomohiro Taniguchi}
 \affiliation{
 National Institute of Advanced Industrial Science and Technology (AIST), Spintronics Research Center, Tsukuba, Ibaraki 305-8568, Japan 
 }

 \begin{abstract}
{ 
A theoretical formalism is developed to simultaneously solve equation of motion of the magnetizations in two ferromagnets 
and the spin-pumping induced spin transport equation. 
Based on the formalism, a coupled motion of the magnetizations in a self-oscillation state is studied. 
The spin pumping is found to induce an in-phase synchronization of the magnetizations for the oscillation around the easy axis. 
For an out-of-plane self-oscillation around the hard axis, on the other hand, 
the spin pumping leads to an in-phase synchronization in a small current region, 
whereas an antiphase synchronization is excited in a large current region. 
An analytical theory based on the phase equation reveals that the phase difference between the magnetizations in a steady state 
depends on the oscillation direction, clockwise or counterclockwise, of the magnetizations. 
}
 \end{abstract}

 \pacs{85.75.-d, 75.78.-n, 05.45.Xt, 72.25.-b}
 \maketitle

% ================================================================================================================================================================================= %

% ===================================================================================================================================================================================== %

\section{Introduction}
\label{sec:Introduction}

Coupled dynamics of magnetizations \cite{tserkovnyak03,tserkovnyak03JAP,heinrich03,smith09,takahashi14,chiba15} in a magnetic multilayer 
via spin pumping \cite{silsbee79,mizukami02a,mizukami02b,tserkovnyak02a,tserkovnyak02b} has provided 
new insight of the relaxation mechanism in nanostructured ferromagnets. 
The spin pumping has been investigated by measuring the linewidth of the power spectrum of ferromagnetic resonance (FMR). 
For example, in a magnetic trilayer consisting of two ferromagnets separated by a thin nonmagnet, 
an appreciable increase in the linewidth was found when the resonance fields of two ferromagnets were well separated, 
whereas a reduction of the linewidth was observed when the resonant fields were close to each other \cite{heinrich03}. 
This is because the in-phase processions of the magnetizations result in the cancellation of the spin pumping. 
A giant enhancement of the damping for an antiphase precession mode was also theoretically predicted in the trilayer system \cite{takahashi14}. 
These results indicate that the phases of the magnetizations, or strictly speaking phase difference, play a key role in the coupled magnetization dynamics via spin pumping. 
%Nevertheless of these extensive previous works, 
%we would like to revisit the problem of the coupled magnetization dynamics via spin pumping due to the following reason. 

% ===================================================================================================================================================================================== %

Recently, the coupled dynamics of the magnetizations has attracted much attention from viewpoints of nonlinear science and practical applications \cite{locatelli14,grollier16,kudo17,torrejon17}. 
An assemblage of spin torque oscillators (STOs) coupled through 
the magnetic \cite{kaka05,mancoff05,urazhdin10,locatelli15,awad17} and/or electric \cite{rippard05,kudo06,nakada12,khalsa15,tsunegi16,taniguchi17} interactions, 
or a forced and self-interacted STO, 
is an interesting example of a synchronized system \cite{strogatz01,kuramoto03,pikovsky03,stankovski17,slavin09}. 
%and is applicable to practical devices such as microwave generators and neuromorphic computing \cite{locatelli14,grollier16,kudo17,torrejon17}. 
In an array of STOs connected to each other proposed in Ref. \cite{kudo17}, for example, the spin pumping may be another and unavoidable mechanism of the coupling 
because the STOs inject spin currents to each other through a common electrode via spin pumping, as in the case of FMR experiments. 
Therefore, we are motivated to investigate what kind of phase dynamics is induced in STOs as a result of the coupling via spin pumping. 
Note that there is an important difference between FMR in a ferromagnetic multilayer and the synchronization of STOs. 
The FMR is a harmonic oscillation of the magnetization excited by an oscillating magnetic field. 
%with a small amplitude excited by an oscillating magnetic field. 
Importantly, in principle, the phases of the magnetizations in a multilayer can be independently controlled by tuning those of the microwaves applied to each layer. 
On the other hand, an STO is a nonlinear oscillator excited by a direct current. 
In a mutual synchronization of STOs, the phases of the oscillators are determined as a result of the interaction among the STOs. 
The phase dynamics in STOs caused by spin pumping however has not been studied yet. 

% ===================================================================================================================================================================================== %

In this paper, we develop a theoretical formalism to simultaneously solve equation of motion of the magnetizations in two ferromagnets 
and the spin transport equation originated from spin pumping. 
Applying the formalism to an in-plane self-oscillation of the STO around the easy axis, 
the spin pumping is found to induce an in-phase synchronization of the magnetizations. 
On the other hand, for an out-of-plane self-oscillation around the hard axis, 
the spin pumping leads to an in-phase synchronization in a small current region, 
whereas an antiphase synchronization is excited in a large current region. 
An analytical theory based on the Landau-Lifshitz-Gilbert (LLG) equation indicates that 
the phase difference stabilized by the spin pumping is related to the oscillation direction of the magnetization. 

% ===================================================================================================================================================================================== %

The paper is organized as follows. 
In Sec. \ref{sec:LLG equations in the presence of spin pumping}, 
we develop a general formalism to solve the LLG equation in the presence of spin pumping. 
Section \ref{sec:Numerical simulation of the LLG equation}
shows the results of the numerical simulation for the phase synchronizations 
oscillating around an easy and hard axis. 
The phase difference in a steady state is discussed in Sec. \ref{sec:Analytical approach} based on an analytical theory of the LLG equation. 
Section \ref{sec:Summary} shows the summary of this work. 

% ===================================================================================================================================================================================== %

% ===================================================================================================================================================================================== %

\section{LLG equations in the presence of spin pumping}
\label{sec:LLG equations in the presence of spin pumping}

In this section, we show a formalism to solve the LLG equation in the presence of spin pumping. 

% ===================================================================================================================================================================================== %

\subsection{The LLG equation}

Let us imagine two ferromagnets F${}_{k}$ ($k=1,2$) connected by a nonmagnet N having the spin diffusion length longer than the distance between the ferromagnets. 
We use the suffixes such as $k$, F${}_{k}$, and N to distinguish the quantities related to the ferromagnet F${}_{k}$, nonmagnet N, and /or their interface F${}_{k}$/N. 
At this point, we will not specify any properties of the ferromagnets, such as the anisotropy and the shape, for generality. 
We denote the unit vector pointing in the magnetization direction of the F${}_{k}$ layer as $\mathbf{m}_{k}$. 
The magnetization dynamics in the F${}_{k}$ layer is described by the LLG equation,
\begin{equation}
  \dot{\mathbf{m}}_{k}
  =
  -\gamma
  \mathbf{m}_{k}
  \times
  \mathbf{H}_{k}
  +
  \alpha_{0}
  \mathbf{m}_{k}
  \times
  \dot{\mathbf{m}}_{k}
  +
  \bm{\tau}_{k}^{\rm STT}
  +
  \bm{\tau}_{k}^{\rm SP},
  \label{eq:LLG}
\end{equation}
where $\gamma$ and $\alpha_{0}$ are the gyromagnetic ratio and the intrinsic Gilbert damping constant, respectively. 
Here, the intrinsic damping is the damping in the absence of the spin pumping. 
The explicit forms of the magnetic field $\mathbf{H}_{k}$ will be given in Sec. \ref{sec:Numerical simulation of the LLG equation} 
when we study the coupled dynamics of the magnetization for several systems. 

% ===================================================================================================================================================================================== %

% ===================================================================================================================================================================================== %

\begin{figure}%[p]
\centerline{\includegraphics[width=1.0\columnwidth]{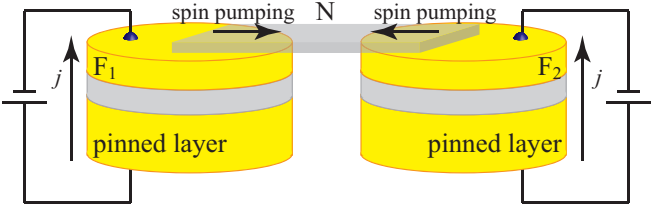}}%\vspace{-3.0ex}
\caption{
         Schematic view of a possible situation in this study. 
         The STO consists of the free (F${}_{k}$) layer, pinned layer, and external voltage. 
         The electric current excites the self-oscillation of the magnetization in the free layer through spin-transfer effect. 
         Simultaneously, the spin currents generated by the spin pumping flow in the nonmagnetic connector N, and lead to a coupled motion of the magnetizations. 
         Note that electric current does not flow in the nonmagnetic connector because the electric potentials at the top surfaces of the free layers are the same. 
         \vspace{-3ex}}
\label{fig:fig1}
\end{figure}

% ===================================================================================================================================================================================== %

% ===================================================================================================================================================================================== %

% ===================================================================================================================================================================================== %

Since the spin pumping occurs as a result of magnetization dynamics, 
an external torque should be applied to the ferromagnet. 
As mentioned in Sec. \ref{sec:Introduction}, we are interested in the coupled dynamics of STOs. 
Therefore, we assume that the magnetization dynamics is excited by a spin torque $\bm{\tau}_{k}^{\rm STT}$. 
Figure \ref{fig:fig1} shows a possible situation, where each STO consists of a free (F${}_{k}$) layer, a pinned layer, and an external voltage. 
The electric current supplied from the voltage excites the spin torque on the magnetization in the free layer. 
The top surfaces of the free layers are connected by the nonmagnet N. 
Since the electric potentials of the free layers at their surfaces are the same, 
we can assume that the electric current does not flow in the nonmagnet N. 
The spin current generated by the spin pumping can, however, flow in the nonmagnet, as in the case of FMR experiments \cite{heinrich03}. 
The spin torque formula is given by \cite{slonczewski96} 
%Therefore, we assume that the magnetization dynamics is excited by a spin torque $\bm{\tau}_{k}^{\rm STT}$ given by \cite{slonczewski96} 
\begin{equation}
  \bm{\tau}_{k}^{\rm STT}
  =
  -\frac{\gamma \hbar \eta j}{2eMd}
  \mathbf{m}_{k}
  \times
  \left(
    \mathbf{p}
    \times
    \mathbf{m}_{k}
  \right),
  \label{eq:STT}
\end{equation}
where $\eta$ is the spin polarization of the current density $j$ injected into the F${}_{k}$ layer. 
The positive current is defined as the electrons flowing from the free to pinned layer. 
%A positive current ($j>0$) corresponds to the electrons flowing from the free to pinned layers. 
The direction of the spin polarization is denoted as $\mathbf{p}$ ($|\mathbf{p}|=1$). 
The saturation magnetization and thickness of the ferromagnet are denoted as $M$ and $d$, respectively. 
%We assume that the electric current does not flow in the nonmagnet connecting the ferromagnets, 
%to avoid any interaction through the current injection, 
%for simplicity. 

% ===================================================================================================================================================================================== %

When the magnetization dynamics is excited by $\bm{\tau}_{k}^{\rm STT}$, 
the spin pumping excites another spin torque given by \cite{tserkovnyak03} 
\begin{equation}
  \bm{\tau}_{k}^{\rm SP}
  =
  \frac{\gamma}{MV}
  \mathbf{m}_{k}
  \times
  \left[
    \left(
      \mathbf{I}_{\rm s}^{{\rm pump}(k)}
      +
      \mathbf{I}_{\rm s}^{{\rm F}_{k}\to{\rm N}}
    \right)
    \times
    \mathbf{m}_{k}
  \right],
  \label{eq:spin_pumping_torque}
\end{equation}
where $\mathbf{I}_{\rm s}^{{\rm pump}(k)}$ is the spin current pumped from the F${}_{k}$ layer \cite{tserkovnyak02a}, 
whereas $-\mathbf{I}_{\rm s}^{{\rm F}_{k}\to{\rm N}}$ is the backflow \cite{tserkovnyak02b,brataas01} originated from the spin accumulation $\bm{\mu}_{\rm N}$ in the nonmagnet. 
The volume of the ferromagnet is denoted as $V$. 
The explicit forms of $\mathbf{I}_{\rm s}^{{\rm pump}(k)}$ and $\mathbf{I}_{\rm s}^{{\rm F}_{k}\to{\rm N}}$ are given by 
\begin{equation}
  \mathbf{I}_{\rm s}^{{\rm pump}(k)}
  =
  \frac{\hbar}{4\pi}
  g_{{\rm r}(k)}
  \mathbf{m}_{k}
  \times
  \dot{\mathbf{m}}_{k},
  \label{eq:spin_pumping}
\end{equation}
\begin{equation}
  \mathbf{I}_{\rm s}^{{\rm F}_{k}\to{\rm N}}
  =
  \frac{-1}{4\pi}
  \left[
    g_{k}^{*}
    \left(
      \mathbf{m}_{k}
      \cdot
      \bm{\mu}_{\rm N}
    \right)
    \mathbf{m}_{k}
    +
    g_{{\rm r}(k)}
    \mathbf{m}_{k}
    \times
    \left(
      \bm{\mu}_{\rm N}
      \times
      \mathbf{m}_{k}
    \right)
  \right],
  \label{eq:spin_current_FN}
\end{equation}
where $g_{{\rm r}(k)}$ is the real part of mixing conductance at the F${}_{k}$/N interface \cite{brataas01}. 
We neglect the imaginary part of the mixing conductance because it is usually much smaller than the real part \cite{zwierzycki05}. 
Another dimensionless conductance $g_{k}^{*}$ is related to the interface resistance $r$ at the F${}_{k}$/N interface, 
as well as the relaxation of the longitudinal spin current inside the ferromagnet; see Appendix \ref{sec:AppendixA}. 
We assume that the spin-relaxation scattering in the nonmagnet is weak and, therefore, the spin current inside the nonmagnetic connector is conserved, i.e., 
\begin{equation}
  \sum_{k=1,2}
  \left[
    \mathbf{I}_{\rm s}^{{\rm pump}(k)}
    +
    \mathbf{I}_{\rm s}^{{\rm F}_{k}\to{\rm N}}
  \right]
  =
  \bm{0}.
  \label{eq:conservation} 
\end{equation}
The spin accumulation in the nonmagnet is determined by solving Eq. (\ref{eq:conservation}) with Eqs. (\ref{eq:spin_pumping}) and (\ref{eq:spin_current_FN}). 
The explicit form of $\bm{\mu}_{\rm N}$ is given by Eq. (\ref{eq:accumulation_sol_tmp}), or equivalently Eq. (\ref{eq:accumulation_sol}), in Appendix \ref{sec:AppendixB}. 
We should note here that $\bm{\mu}_{\rm N}$ is spatially uniform due to the assumption of the weak spin-relaxation scattering in the nonmagnet.
Substituting Eq. (\ref{eq:conservation}) into Eq. (\ref{eq:spin_pumping_torque}), 
we can calculate the spin torque $\bm{\tau}_{k}^{\rm SP}$ due to the spin pumping. 

% ===================================================================================================================================================================================== %

Solving Eq. (\ref{eq:conservation}) with Eqs. (\ref{eq:spin_pumping}) and (\ref{eq:spin_current_FN}) 
with respect to $\bm{\mu}_{\rm N}$, 
and substituting its solution to Eq. (\ref{eq:spin_pumping_torque}), Eq. (\ref{eq:LLG}) becomes 
\begin{equation}
  \mathsf{L}
  \begin{pmatrix}
    \dot{\mathbf{m}}_{1} \\
    \dot{\mathbf{m}}_{2}
  \end{pmatrix}
  =
  \begin{pmatrix}
    -\gamma \mathbf{m}_{1} \times \mathbf{H}_{1} + \bm{\tau}_{1}^{\rm STT} \\
    -\gamma \mathbf{m}_{2} \times \mathbf{H}_{2} + \bm{\tau}_{2}^{\rm STT}
  \end{pmatrix}, 
  \label{eq:LLG_SP}
\end{equation}
where $\mathsf{L}$ is a $6 \times 6$ matrix. 
The components of $\mathsf{L}$ depend on the conductances $g_{k}^{*}$ and $g_{{\rm r}(k)}$ characterizing the amount of spin current in the nonmagnetic connector 
and are explicitly given in Appendix \ref{sec:AppendixB} for general systems. 
When the material parameters of two ferromagnets are identical, the matrix $\mathsf{L}$ is given by 
\begin{equation}
\begin{split}
  \mathsf{L}
  =&
  \hat{I}
  +
  \alpha_{0}
  \begin{pmatrix}
    \mathsf{M}_{1} & \bm{0}_{3} \\
    \bm{0}_{3} & \mathsf{M}_{2}
  \end{pmatrix}
  +
  \alpha^{\prime}
  \begin{pmatrix}
    \mathsf{M}_{1} & \bm{0}_{3} \\
    \bm{0}_{3} & \mathsf{M}_{2}
  \end{pmatrix}
\\
  &+
  \alpha^{\prime}
  \begin{pmatrix}
    \mathsf{N}_{(1)}-\mathsf{N}_{(1,1)}^{\prime} & \mathsf{N}_{(2)}-\mathsf{N}_{(1,2)}^{\prime} \\
    \mathsf{N}_{(1)}-\mathsf{N}_{(2,1)}^{\prime} & \mathsf{N}_{(2)}-\mathsf{N}_{(2,2)}^{\prime}
  \end{pmatrix}
  \label{eq:matrix_L}
\end{split}
\end{equation}
where $\hat{I}$ is the $6 \times 6$ unit matrix, 
and $\alpha^{\prime}$ is defined as 
\begin{equation}
  \alpha^{\prime}
  =
  \frac{\gamma \hbar g_{\rm r}}{4\pi MV}. 
  \label{eq:alpha_prime}
\end{equation}
A $3\times 3$ matrix $\mathsf{M}_{k}$ ($k=1,2$) is defined as 
\begin{equation}
  \mathsf{M}_{k}
  =
  \begin{pmatrix}
    0 & m_{kz} & -m_{ky} \\
    -m_{kz} & 0 & m_{kx} \\
    m_{ky} & -m_{kx} & 0
  \end{pmatrix}, 
\end{equation}
whereas $\bm{0}_{3}$ is the $3\times 3$ zero matrix. 
The first and second terms on the right hand side of Eq. (\ref{eq:matrix_L}) correspond to the term 
$1-\alpha_{0} \mathbf{m}_{k} \times$ in Eq. (\ref{eq:LLG}). 
The third term in Eq. (\ref{eq:matrix_L}) corresponds to the term related to $\mathbf{I}_{\rm s}^{{\rm pump}(k)}$ in Eq. (\ref{eq:spin_pumping_torque}) 
and is the enhancement of the Gilbert damping constant due to the spin pumping for a single ferromagnet \cite{tserkovnyak02a}. 
On the other hand, the last term in Eq. (\ref{eq:matrix_L}) comes from the term related to $\mathbf{I}_{\rm s}^{{\rm F}_{k}\to{\rm N}}$ in Eq. (\ref{eq:spin_pumping_torque}). 
The $(a,b)$ ($a,b=1,2,3$ or $x$,$y$,$z$) components of $3 \times 3$ matrices $\mathsf{N}_{(\ell)}$ and $\mathsf{N}_{(k,\ell)}^{\prime}$ are given by 
\begin{equation}
  N_{(\ell)ab}
  =
  g_{\rm r}
  \epsilon_{bcd}
  K_{ac}^{-1}
  m_{\ell d},
  \label{eq:matrix_N_component}
\end{equation}
\begin{equation}
  N_{(k,\ell)ab}^{\prime}
  =
  \sum_{c=x,y,z}
  m_{kc}
  m_{ka}
  N_{(\ell)cb},
  \label{eq:matrix_Np_component}
\end{equation}
where $\epsilon_{abc}$ is the Levi-Civita asymmetric symbol ($\epsilon_{123}$ or $\epsilon_{xyz}=+1$), 
whereas $K_{ab}^{-1}$ is the $(a,b)$ component of a $3\times 3$ matrix $\mathsf{K}^{-1}$ given by 
(see also Appendix \ref{sec:AppendixB}) 
\begin{equation}
\begin{split}
  K_{ab}^{-1}
  =&
  \frac{(1-\nu^{2})\delta_{ab}+(1+\nu)\nu(m_{1a}m_{1b}+m_{2a}m_{2b})}{2g_{\rm r}[1-\nu^{2}(\mathbf{m}_{1}\cdot\mathbf{m}_{2})^{2}]}
\\
  &+
  \frac{\nu^{2} \mathbf{e}_{a}\cdot(\mathbf{m}_{1}\times\mathbf{m}_{2})\mathbf{e}_{b}\cdot(\mathbf{m}_{1}\times\mathbf{m}_{2})}{2g_{\rm r}[1-\nu^{2}(\mathbf{m}_{1}\cdot\mathbf{m}_{2})^{2}]}.
  \label{eq:matrix_K_inv_component}
\end{split}
\end{equation}
Here, $\mathbf{e}_{a}$ ($a=x,y,z$) is the unit vector pointing in the $a$ direction, 
whereas $\nu=(g_{\rm r}-g^{*})/(g_{\rm r}+g^{*})$ \cite{tserkovnyak03}. 
Note that the off-diagonal components, $\mathsf{N}_{(2)}-\mathsf{N}_{(1,2)}^{\prime}$ and $\mathsf{N}_{(1)}-\mathsf{N}_{(2,1)}^{\prime}$ in Eq. (\ref{eq:matrix_L}), 
lead to a coupled motion of the magnetizations. 
In Sec. \ref{sec:Numerical simulation of the LLG equation}, we solve Eq. (\ref{eq:LLG_SP}) for the coupled STOs to investigate 
the role of spin pumping on the phase dynamics of the magnetizations. 

% ===================================================================================================================================================================================== %

% ===================================================================================================================================================================================== %

\subsection{Approximated formula of torque due to spin pumping}

We should note here that the explicit forms of $\bm{\tau}_{k}^{\rm SP}$ have been obtained for specific cases \cite{tserkovnyak03,takahashi14,taniguchi07,taniguchi14}. 
For example, when magnetizations oscillate around a common $z$ axis with a small amplitude 
and $g_{\rm r(1)}=g_{\rm r(2)}$, 
$m_{ka}m_{kb}$ ($a,b=x,y$) in Eq. (\ref{eq:matrix_K_inv_component}) is the higher order term of the small amplitude, 
and $\mathbf{m}_{1}\times\mathbf{m}_{2} \simeq \bm{0}$. 
This means that the off-diagonal components of $K_{ab}$ are negligible 
and $K_{11}^{-1}=K_{22}^{-1}\simeq 1/(2 g_{\rm r})$. 
Then, we find that 
\begin{equation}
  \bm{\mu}_{\rm N}
  \simeq
  \frac{\hbar}{2}
  \left(
    \mathbf{m}_{1}
    \times
    \dot{\mathbf{m}}_{1}
    +
    \mathbf{m}_{2}
    \times
    \dot{\mathbf{m}}_{2}
  \right).
  \label{eq:spin_accumulation_approx}
\end{equation}
Since $\mathbf{m}_{1},\mathbf{m}_{2} \simeq +\mathbf{e}_{z}$, 
we can also assume that $\mathbf{m}_{k}\cdot\bm{\mu}_{\rm N} \simeq 0$ up to the first orders of $m_{kx}$ and $m_{ky}$. 
%This means the terms related to $\mathsf{N}_{(k,\ell)}^{\prime}$ can be neglected from Eq. (\ref{eq:matrix_L}). 
We then notice that 
$\bm{\tau}_{k}^{\rm SP}$ is approximated as 
\begin{equation}
  \bm{\tau}_{k}^{\rm SP}
  \simeq
  \frac{\alpha^{\prime}}{2}
  \left(
    \mathbf{m}_{k}
    \times
    \dot{\mathbf{m}}_{k}
    -
    \mathbf{m}_{k^{\prime}}
    \times
    \dot{\mathbf{m}}_{k^{\prime}}
  \right), 
  \label{eq:tau_SP_tserkovnyak}
\end{equation}
where $(k,k^{\prime})=(1,2)$ or $(2,1)$. 
Equation (\ref{eq:tau_SP_tserkovnyak}) is the spin torque due to the spin pumping obtained in Ref. \cite{tserkovnyak03JAP}. 
The first term on the right hand side of Eq. (\ref{eq:tau_SP_tserkovnyak}) can be regarded as an enhancement of the Gilbert damping constant. 
On the other hand, the second term induces the coupled motion of the magnetizations. 

% ===================================================================================================================================================================================== %

We emphasize that these analytical formulas are obtained by assuming specific alignments of the magnetizations. 
For example, Eq. (\ref{eq:tau_SP_tserkovnyak}) is valid only for the small amplitude oscillations around a common axis. 
It was shown for a different situation that the spin pumping affects not only the damping but also the frequency \cite{taniguchi07}, 
implying that $\bm{\tau}_{k}^{\rm SP}$ has a projection to the direction of $\dot{\mathbf{m}}_{k}$. 
On the other hand, we are interested in the magnetization alignment after the spin pumping has induced the coupled motion of magnetizations. 
Therefore, we do not use any assumption of the magnetization alignment nor analytical expression of $\bm{\tau}_{k}^{\rm SP}$ in the numerical simulation shown in Sec. \ref{sec:Numerical simulation of the LLG equation}. 
%Rather, we developed a formalism to solve Eq. (\ref{eq:conservation}) with the equation of the motion of the LLG equation self-consistently. 
We, simultaneously however, note that Eq. (\ref{eq:tau_SP_tserkovnyak}) is useful to understand 
analytically the phase synchronization shown below; see Sec \ref{sec:Analytical approach}. 

% ===================================================================================================================================================================================== %

We also note that Eq. (\ref{eq:tau_SP_tserkovnyak}) is an approximated solution of the torque when $\mathbf{m}_{1},\mathbf{m}_{2} \simeq +\mathbf{e}_{z}$, 
and must not be used directly in the numerical simulation of the LLG equation. 
The LLG equation assumes that the norm of the magnetization, $|\mathbf{m}_{k}|=1$, is conserved, 
indicating that all torques in the LLG equation should be orthogonal to the direction of $\mathbf{m}_{k}$. 
In general, however, the second term of Eq. (\ref{eq:tau_SP_tserkovnyak}) has a finite projection to the direction of $\mathbf{m}_{k}$, i.e., 
$\mathbf{m}_{k}\cdot(\mathbf{m}_{k^{\prime}}\times\dot{\mathbf{m}}_{k^{\prime}}) \neq 0$, 
and as a result, the norm of the magnetization is not conserved in Eq. (\ref{eq:tau_SP_tserkovnyak}). 
Therefore, even for the situation in which Eq. (\ref{eq:spin_accumulation_approx}) is valid, 
the second term on the right hand side of Eq. (\ref{eq:tau_SP_tserkovnyak}) should be replaced by 
$-(\alpha^{\prime}/2) \mathbf{m}_{k} \times [(\mathbf{m}_{k^{\prime}} \times \dot{\mathbf{m}}_{k^{\prime}}) \times \mathbf{m}_{k} ]$ 
when it is applied to the numerical simulation. 

% ===================================================================================================================================================================================== %

% ===================================================================================================================================================================================== %

% ===================================================================================================================================================================================== %

\section{Numerical simulation of the LLG equation}
\label{sec:Numerical simulation of the LLG equation}

In this section, we solve Eq. (\ref{eq:LLG_SP}) numerically and investigate the coupled motion of the magnetizations. 
We should note that the STOs are typically classified into two types, 
where the magnetization oscillates around its easy \cite{liu12} or hard axis \cite{houssameddine07}. 
Therefore, we study the phase synchronization for these two cases. 

We assume that the material parameters of two STOs are identical, for simplicity. 
The values of the parameters are derived from typical ferromagnets, such as NiFe \cite{fert99} and CoFeB \cite{kim16}, used in STOs 
and the first-principle calculations \cite{zwierzycki05} (see also Appendix \ref{sec:AppendixA}): 
$\rho=300$ $\Omega$nm, 
$\beta=0.75$, 
$\lambda_{\rm sd}=5.0$ nm, 
$r=0.25$ k$\Omega$nm${}^{2}$, and 
$p_{g}=0.50$. 
The thickness of the ferromagnet is $d=2.0$ nm. 
Then, $g^{*}/S$ given by Eq. (\ref{eq:g_star}) is 1.38 nm${}^{-2}$, 
where $S$ is the cross section area of the F/N interface. 
On the other hand, the real part of the mixing conductance is set to be $g_{\rm r}/S=15$ nm${}^{-2}$. 
Then, $\nu$ is calculated as $0.83$. 
The saturation magnetization $M$, the gyromagnetic ratio $\gamma$, and the intrinsic Gilbert damping constant $\alpha_{0}$ are 
assumed as $1500$ emu/c.c., $1.764 \times 10^{7}$ rad/(Oe s), and $0.010$, respectively. 
Then, $\alpha^{\prime}$ becomes $0.0074$. 
The spin polarization $\eta$ is set to be 0.50, 
whereas the in-plane anisotropy field $H_{\rm K}$ for the oscillation around the easy axis is $200$ Oe. 

The results shown in the main text are obtained for the initial conditions of 
$\mathbf{m}_{1}(0)=(\cos 5^{\circ},\sin 5^{\circ},0)$ and $\mathbf{m}_{2}(0)=(\cos 10^{\circ},\sin 10^{\circ},0)$ 
for both the oscillations around the easy and hard axes. 
The phase difference in the steady state is, however, independent of the initial states. 
The role of the angular dependence of the spin torque is discussed in Appendix \ref{sec:AppendixC}. 

% ===================================================================================================================================================================================== %

\subsection{Oscillation around easy axis}

% ===================================================================================================================================================================================== %

% ===================================================================================================================================================================================== %

\begin{figure*}%[p]
\centerline{\includegraphics[width=2.0\columnwidth]{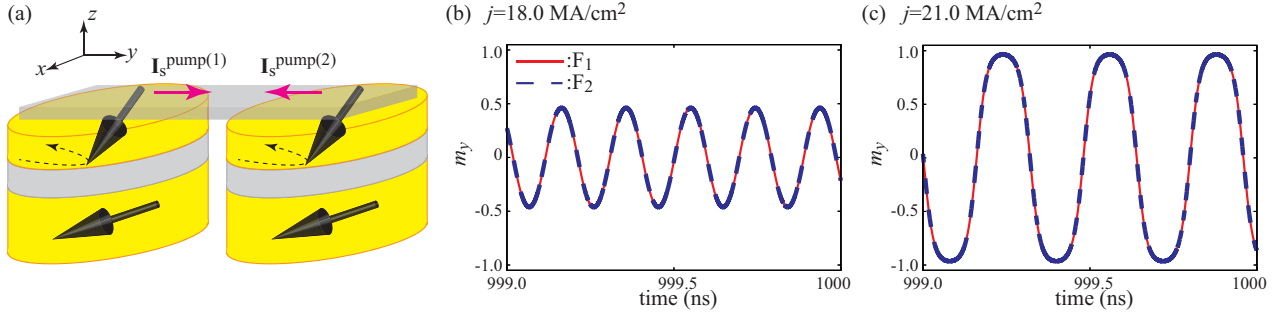}}%\vspace{-3.0ex}
\caption{
        (a) Schematic view of the self-oscillation excited in an in-plane magnetized ferromagnet. 
            The magnetizations oscillate around the easy axis along the $x$ direction. 
            When these ferromagnets are connected by a nonmagnet, the spin currents due to spin pumping are injected into each other. 
        Examples of the oscillations of $m_{ky}$ ($k=1,2$) for (b) small ($j=18.0$ MA/cm${}^{2}$) and (c) large ($21.0$ MA/cm${}^{2}$) currents, 
        where the red solid (blue dotted) line corresponds to $m_{1y}$ ($m_{2y}$). 
         \vspace{-3ex}}
\label{fig:fig2}
\end{figure*}

% ===================================================================================================================================================================================== %

% ===================================================================================================================================================================================== %

First, let us study the coupled dynamics in the STOs in which the magnetizations oscillate around the easy axis. 
Figure \ref{fig:fig2}(a) is a schematic view of the system, 
where the free layers of two STOs are connected by the nonmagnet 
and interact each other via spin pumping. 
We assume an in-plane magnetized free layer, as in the case of an experiment in Ref. \cite{liu12}, 
where the magnetic field consists of the in-plane anisotropy field $H_{\rm K}$ 
and the shape anisotropy field along the perpendicular direction, i.e., 
\begin{equation}
  \mathbf{H}_{k}
  =
  H_{\rm K}
  m_{kx}
  \mathbf{e}_{x}
  -
  4\pi M m_{kz}
  \mathbf{e}_{z},
  \label{eq:field_in_plane}
\end{equation}
where the in-plane easy axis is parallel to the $x$ axis. 
We note that this type of the free layer has two energetically stable states at $\mathbf{m}_{k}=\pm\mathbf{e}_{x}$. 
For the sake of convention, we assume that the magnetizations initially locate near $\mathbf{m}_{k}=+\mathbf{e}_{x}$. 
The spin polarization is parallel to the easy axis direction, $\mathbf{p}=+\mathbf{e}_{x}$. 
In the absence of the coupling, this type of free layer shows self-oscillation when the current density $j$ is in the range of 
$J_{\rm c}<j<J^{*}$, where $J_{c}$ and $J^{*}$ are given by \cite{bazaliy07,taniguchi13PRB} 
\begin{equation}
  J_{\rm c}
  =
  \frac{2 \alpha_{0}eMd}{\hbar \eta}
  \left(
    H_{\rm K}
    +
    2\pi M
  \right),
  \label{eq:Jc}
\end{equation}
\begin{equation}
  J^{*}
  =
  \frac{4 \alpha_{0} eMd}{\pi \hbar \eta}
  \sqrt{
    4\pi M 
    \left(
      H_{\rm K}
      +
      4\pi M
    \right)
  }. 
  \label{eq:J_star}
\end{equation}
The values of $J_{\rm c}$ and $J^{*}$ in the present calculations are $17.5$ and $22.0$ MA/cm${}^{2}$, respectively. 
In this type of STO, the oscillation orbit around the easy axis becomes large with increasing current magnitude. 
Figures \ref{fig:fig2}(b) and \ref{fig:fig2}(c) show the dynamics of $m_{ky}(t)$ ($k=1,2$) 
for relatively small ($j=18.0$ MA/cm${}^{2}$ $\simeq J_{\rm c}$) and large ($j=21.0$ MA/cm${}^{2}$ $\simeq J^{*}$) current regions. 
Starting from different initial conditions, we find that $\mathbf{m}_{1}$ and $\mathbf{m}_{2}$ finally show an in-phase synchronization 
[$\mathbf{m}_{1}(t)=\mathbf{m}_{2}(t)$] for both the small and large current regions. 

% ===================================================================================================================================================================================== %

% ===================================================================================================================================================================================== %

\subsection{Oscillation around hard axis}

% ===================================================================================================================================================================================== %

% ===================================================================================================================================================================================== %

\begin{figure*}%[p]
\centerline{\includegraphics[width=2.0\columnwidth]{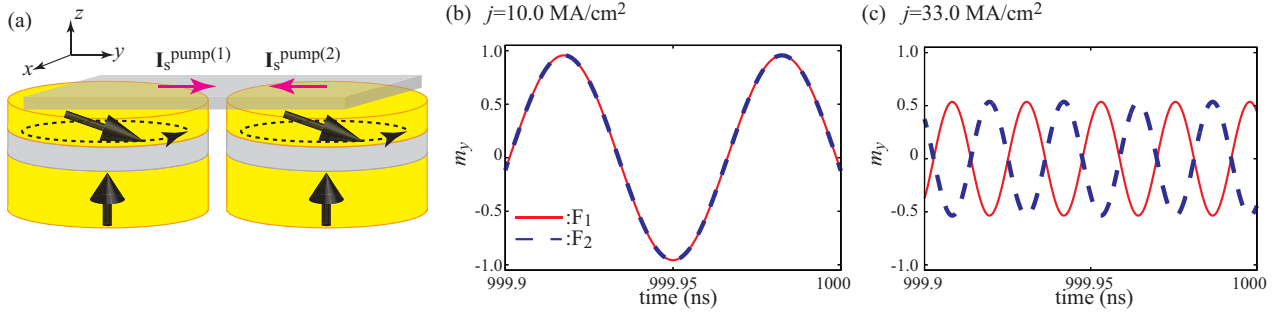}}%\vspace{-3.0ex}
\caption{
        (a) Schematic view of an out-of-plane self-oscillation of the magnetization. 
            The magnetizations oscillate around the hard axis along the $z$ direction. 
        Examples of the oscillations of $m_{ky}$ ($k=1,2$) for (b) small ($j=10.0$ MA/cm${}^{2}$) and (c) large ($33.0$ MA/cm${}^{2}$) currents, 
        where the red solid (blue dotted) line corresponds to $m_{1y}$ ($m_{2y}$). 
        Note that the time scales in the horizontal axes of (b) and (c) are different from those in Fig. \ref{fig:fig2}. 
         \vspace{-3ex}}
\label{fig:fig3}
\end{figure*}

% ===================================================================================================================================================================================== %

% ===================================================================================================================================================================================== %

Next, let us consider the coupled dynamics in the STOs when the magnetizations oscillate around the hard axis. 
Figure \ref{fig:fig3}(a) is a schematic view of the system. 
The STOs show out-of-plane oscillations, as in the case of an experiment in Ref. \cite{houssameddine07}. 
The magnetic field consists of the shape anisotropy field along the perpendicular axis, 
\begin{equation}
  \mathbf{H}_{k}
  =
  -4\pi M 
  m_{kz}
  \mathbf{e}_{z}.
  \label{eq:field_out_of_plane}
\end{equation}
while $\mathbf{p}=+\mathbf{e}_{z}$. 
The positive (negative) current moves the magnetization toward the negative (positive) $z$ direction 
and excites an out-of-plane self-oscillation around the $z$ axis. 
In the absence of the coupling via spin pumping, the self-oscillation appears when the current density is in the range of \cite{taniguchi16} 
\begin{equation}
  0
  <
  |j|
  <
  \frac{2 \alpha_{0}eMd}{\hbar \eta}
  4\pi M, 
\end{equation}
where the upper boundary is approximately $[2 \alpha_{0}eMd/(\hbar \eta)]4\pi M \simeq 34.4$ MA/cm${}^{2}$. 
In this type of STO, the spin torque tries to switch the magnetization to the direction of the $z$ axis. 
As a result, the oscillation amplitude becomes small by increasing the current magnitude. 
Figures \ref{fig:fig3}(b) and \ref{fig:fig3}(c) show the dynamics of $m_{ky}(t)$ 
in a relatively small ($j=10.0$ MA/cm${}^{2}$) and large ($j=33.0$ MA/cm${}^{2}$) current regions. 
It is found that the in-phase synchronization is excited in the small current region, 
whereas an antiphase synchronization appears in the large current region. 

% ===================================================================================================================================================================================== %

% ===================================================================================================================================================================================== %

\begin{figure}%[p]
\centerline{\includegraphics[width=1.0\columnwidth]{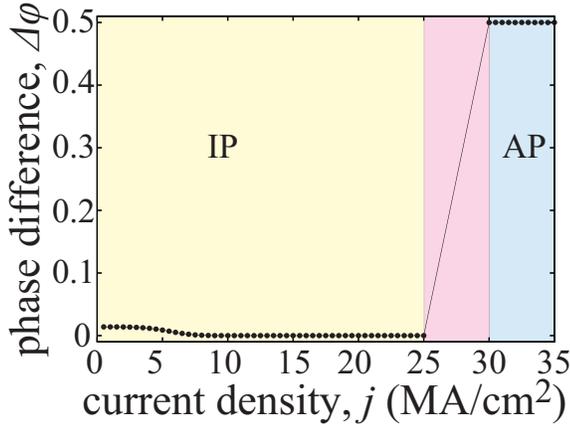}}%\vspace{-3.0ex}
\caption{
        Dependence of the phase difference $\Delta\varphi$ between the magnetizations in a steady state on the current density $j$ for the out-of-plane self-oscillation. 
        The in-phase (IP, yellow) and antiphase (AP, blue) states correspond to $\Delta\varphi=0$ and $\Delta\varphi=0.5$, respectively. 
        The phase difference in the intermediate region, $25 \lesssim j \lesssim 30$ MA/cm${}^{2}$, shown by pink, is not well-defined. 
         \vspace{-3ex}}
\label{fig:fig4}
\end{figure}

% ===================================================================================================================================================================================== %

% ===================================================================================================================================================================================== %

Figure \ref{fig:fig4} summarizes the current dependence of the phase difference ($\Delta\varphi$) between the STOs showing the out-of-plane oscillations, 
where $\Delta\varphi=0$ and $\Delta\varphi=0.5$ correspond to the in-phase and antiphase, respectively \cite{taniguchi17}. 
For the present system, the in-phase synchronization appears for $0<|j| \lesssim 25$ MA/cm${}^{2}$, 
whereas the antiphase synchronization is excited for $|j| \gtrsim 30$ MA/cm${}^{2}$. 
We should emphasize here that these phase differences are stable, 
i.e., once the phase difference is saturated to one of these values, it does not change any more. 
On the other hand, the phase difference which appears in the intermediate region, $25 \lesssim |j| \lesssim 30$ MA/cm${}^{2}$, is not well defined due to the following reason. 

% ===================================================================================================================================================================================== %

% ===================================================================================================================================================================================== %

\begin{figure}%[p]
\centerline{\includegraphics[width=1.0\columnwidth]{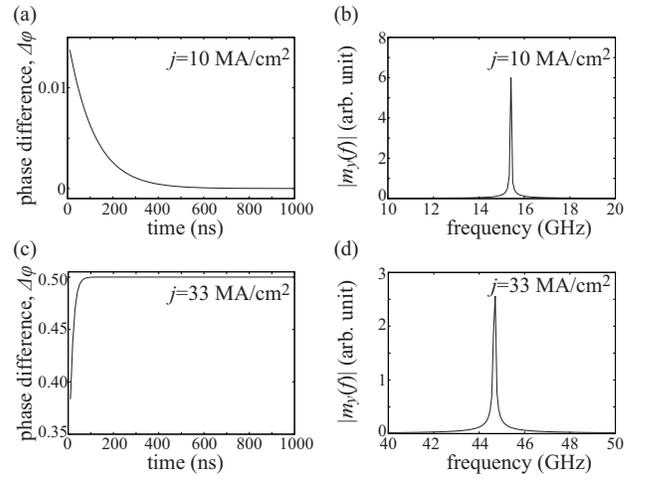}}%\vspace{-3.0ex}
\caption{
        (a) Time evolutions of $\Delta\varphi$ for $j=10.0$ MA/cm${}^{2}$ 
        and (b) the Fourier transformation of $m_{1y}(t)$ for the same current. 
        (c) The time evolutions of $\Delta\varphi$ for $j=33.0$ MA/cm${}^{2}$ 
        and (d) the Fourier transformation of $m_{1y}(t)$ for the same current. 
         \vspace{-3ex}}
\label{fig:fig5}
\end{figure}

% ===================================================================================================================================================================================== %

% ===================================================================================================================================================================================== %

% ===================================================================================================================================================================================== %

% ===================================================================================================================================================================================== %

\begin{figure}%[p]
\centerline{\includegraphics[width=1.0\columnwidth]{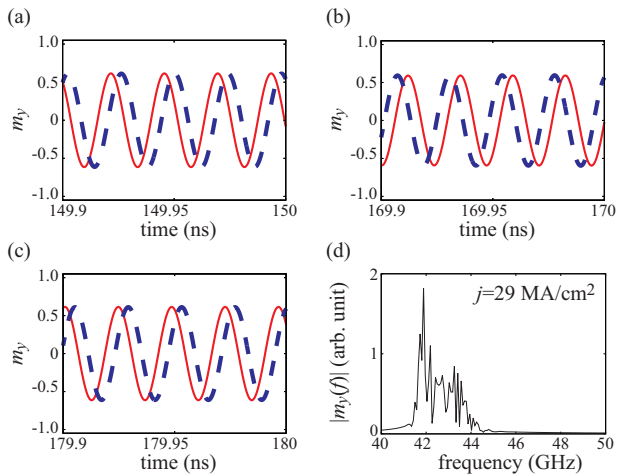}}%\vspace{-3.0ex}
\caption{
        (a)-(c) Oscillations of $m_{1y}$ and $m_{2y}$ at the different time ranges 
        and (d) the Fourier transformation of $m_{1y}$ for $j=29.0$ MA/cm${}^{2}$. 
        The red solid (blue dotted) line in (a)-(c) corresponds to $m_{1y}$ ($m_{2y}$). 
         \vspace{-3ex}}
\label{fig:fig6}
\end{figure}

% ===================================================================================================================================================================================== %

% ===================================================================================================================================================================================== %

To define the phase difference, the oscillators should oscillate with the same frequency \cite{strogatz01,kuramoto03,pikovsky03}. 
Figure \ref{fig:fig5}(a) shows the time evolutions of the phase difference $\Delta\varphi$ for $j=10.0$ MA/cm${}^{2}$. 
It can be seen from the figure that the phase differences monotonically saturate to a stable value (in-phase, $\Delta\varphi=0$). 
The Fourier spectrum of the oscillating component [$m_{y}(t)$] after the phase difference is fixed shows a sharp peak 
at a certain frequency, as shown in Fig. \ref{fig:fig5}(b). 
The same results are obtained for the other stable (antiphase) state. 
As shown in Fig. \ref{fig:fig5}(c), the phase difference for $j=33.0$ MA/cm${}^{2}$ monotonically saturates to a stable value. 
The Fourier spectrum for this current also shows a sharp peak at the oscillation frequency, as shown in Fig. \ref{fig:fig5}(d). 
We emphasize that the phase differences in these cases are well defined because two magnetizations always oscillate with the same single frequency. 
On the other hand, at the intermediate region, the oscillation frequency of the magnetizations is not fixed to a certain value. 
Figures \ref{fig:fig6}(a)-\ref{fig:fig6}(c) show the oscillations of $m_{1y}(t)$ and $m_{2y}(t)$ in the intermediate region, 
where the current density $j$ is $29.0$ MA/cm${}^{2}$. 
For example, for the time range of $149.9 \le t \le 150$ ns, 
the phase of $m_{1y}$ is preceded by that of $m_{2y}$. 
For $169.9 \le t \le 170$ ns, however, the sequential order of $m_{1y}$ and $m_{2y}$ is inversed. 
%$m_{1y}$ is forwarded to $m_{2y}$. 
Further more, when $179.9 \le t \le 180$, $m_{1y}$ is again preceded by $m_{2y}$. 
These results indicate that the oscillation frequencies of $m_{1y}$ and $m_{2y}$ are not locked. 
This argument is also confirmed from the fact that the Fourier spectrum has several peaks, as shown in Fig. \ref{fig:fig6}(d). 
Therefore, the phase difference in the intermediate region is not well defined. 

% ===================================================================================================================================================================================== %

% ===================================================================================================================================================================================== %

\section{Analytical approach}
\label{sec:Analytical approach}

In this section, we show the derivation of the phase equation from the LLG equation 
and discuss analytically the phase difference. 

% ===================================================================================================================================================================================== %

\subsection{LLG equation in spherical coordinate}

We start by showing the LLG equation in the absence of the spin pumping. 
Since we are interested in the phase dynamics of the magnetizations, 
it is convenient to describe the magnetization dynamics in terms of the phase. 
Therefore, by using spherical coordinates, we introduce the zenith and azimuth angles $(\theta_{k},\varphi_{k})$ as 
$\mathbf{m}_{k}=(\sin\theta_{k}\cos\varphi_{k},\sin\theta_{k}\sin\varphi_{k},\cos\theta_{k})$. 
The LLG equation in the absence of the spin pumping is given by 
\begin{equation}
  \frac{d \theta_{k}}{dt}
  =
  -\frac{\gamma}{M \sin\theta_{k}}
  \frac{\partial E}{\partial \varphi_{k}}
  -
  \frac{\gamma \hbar \eta j}{2eMd}
  \frac{\partial}{\partial \theta_{k}}
  \mathbf{m}_{k}
  \cdot
  \mathbf{p}
  -
  \alpha_{0}
  \sin\theta_{k}
  \frac{d \varphi_{k}}{dt},
  \label{eq:LLG_theta}
\end{equation}
\begin{equation}
  \sin\theta_{k}
  \frac{d \varphi_{k}}{dt}
  =
  \frac{\gamma}{M}
  \frac{\partial E}{\partial \theta_{k}}
  -
  \frac{1}{\sin\theta_{k}}
  \frac{\gamma \hbar \eta j}{2eMd}
  \frac{\partial}{\partial\varphi_{k}}
  \mathbf{m}_{k}
  \cdot
  \mathbf{p}
  +
  \alpha_{0}
  \frac{d \theta_{k}}{dt}, 
  \label{eq:LLG_varphi}
\end{equation}
where $E$ is the magnetic energy density of the ferromagnet related to the magnetic field $\mathbf{H}$ via $E=-M \int d \mathbf{m}\cdot\mathbf{H}$. 
The first terms of Eqs. (\ref{eq:LLG_theta}) and (\ref{eq:LLG_varphi}) determine the oscillation frequency of the magnetization in the self-oscillation state. 
On the other hand, the second and third terms, which are the spin torque by the electric current and the intrinsic damping torque, 
cancel each other in the self-oscillation state to sustain the oscillation \cite{bertotti09}. 

% ===================================================================================================================================================================================== %

It is necessary to find the relation between the phase of the oscillator commonly used in nonlinear science \cite{strogatz01,kuramoto03,pikovsky03,stankovski17} 
and the angles $(\theta_{k},\varphi_{k})$ describing the magnetization dynamics. 
It should be emphasized that the choice of the direction of the axis in the Cartesian coordinate here is arbitrary. 
Let us consider a small-amplitude oscillation of the magnetization 
and define a Cartesian coordinate $XYZ$, where the $Z$ axis is parallel to the precession axis of the magnetization. 
For example, the $Z$ axis corresponds to the $x$ axis for the oscillation around the easy axis shown in Fig. \ref{fig:fig2}(a), 
whereas the $Z$ axis is the $z$ axis for the oscillation around the hard axis shown in Fig. \ref{fig:fig3}(a). 
The zenith angle $\theta_{k}$ is a tilted angle of the magnetization from the precession axis. 
Then, in the small-amplitude limit, $\theta_{k} (\to 0,\pi)$ can be regarded as a constant, 
whereas $\varphi_{k}$ can be regarded as the phase in the oscillators \cite{strogatz01,kuramoto03,pikovsky03,stankovski17}. 
%This is because the small-amplitude oscillation can be well described by the trigonometric functions with a constant frequency. 
In this manner, we can study the phase difference in a steady state from Eq. (\ref{eq:LLG_varphi}). 
The small-amplitude assumption works relatively well to the oscillation around the in-plane easy axis 
because the large demagnetization field suppresses the oscillation amplitude. 
On the other hand, for the out-of-plane oscillation, it is applicable only for the oscillation near the $z$ axis, which corresponds to the large current limit.

\subsection{Phase equation in the presence of spin pumping}

In this section, we investigate the phase equation for both the in-plane and out-of-plane oscillations. 
We use Eq. (\ref{eq:tau_SP_tserkovnyak}) as an approximated formula of the spin torque due to spin pumping 
because we focus on the small amplitude limit. 

% ===================================================================================================================================================================================== %

% ===================================================================================================================================================================================== %

First, we consider the oscillation around the easy axis, as shown in Fig. \ref{fig:fig2}(a). 
Therefore, the $Z$ axis mentioned above corresponds to the $x$ axis in Fig. \ref{fig:fig2}(a). 
%In the case of the in-plane magnetized system, 
%the oscillation trajectory is described by the elliptic function \cite{bertotti09}. 
%Therefore, as mentioned above, let us apply the small-amplitude assumption. 
Since we are interested in the role of the coupling on the phase, 
let us focus on the coupling term in the following discussion. 
We note that %, under the assumption of the small-amplitude oscillation, 
the second term of Eq. (\ref{eq:tau_SP_tserkovnyak}) becomes 
\begin{equation}
\begin{split}
  &
  -\frac{\alpha^{\prime}}{2}
  \mathbf{m}_{k^{\prime}}
  \times
  \dot{\mathbf{m}}_{k^{\prime}}
%  &\simeq
%  -\frac{\alpha^{\prime}}{2}
%  \begin{pmatrix}
%    \sin\theta_{k^{\prime}}\cos\varphi_{k^{\prime}} \\
%    \sin\theta_{k^{\prime}}\sin\varphi_{k^{\prime}} \\
%    \cos\theta_{k^{\prime}}
%  \end{pmatrix}
%  \times
%  \begin{pmatrix}
%    -\dot{\varphi}_{k^{\prime}} \sin\theta_{k^{\prime}}\sin\varphi_{k^{\prime}} \\
%    \dot{\varphi}_{k^{\prime}} \sin\theta_{k^{\prime}}\cos\varphi_{k^{\prime}} \\
%    0
%  \end{pmatrix}
  \simeq
  \frac{\alpha^{\prime}}{2}
  \dot{\varphi}_{k^{\prime}}
  \begin{pmatrix}
    \sin\theta_{k^{\prime}}\cos\theta_{k^{\prime}}\cos\varphi_{k^{\prime}} \\
    \sin\theta_{k^{\prime}}\cos\theta_{k^{\prime}}\sin\varphi_{k^{\prime}} \\
    0
  \end{pmatrix}. 
\end{split}
\end{equation}
Then, adding Eq. (\ref{eq:tau_SP_tserkovnyak}) to Eq. (\ref{eq:LLG_varphi}), 
the phase $\varphi_{k}$ of the magnetization $\mathbf{m}_{k}$ obeys the following equation, 
\begin{equation}
  \frac{d \varphi_{k}}{dt}
  \sim
  \omega
  -
  \frac{\alpha^{\prime}}{2}
  \dot{\varphi}_{k^{\prime}}
  \cos\theta_{k^{\prime}}
  \sin(\varphi_{k}-\varphi_{k^{\prime}}), 
  \label{eq:LLG_varphi_tmp}
\end{equation}
where $\omega$ is the angular velocity of the oscillation. 
Let us define the phase difference as $\Delta\varphi=\varphi_{k}-\varphi_{k^{\prime}}$. 
%The small amplitude oscillation means that $\theta_{k}^{\prime} \simeq 0,\pi$. 
%We note that the direction of the magnetization oscillation is determined by the torque due to the field, $-\gamma \mathbf{m}\times\mathbf{H}$. 
%In the present approximation, $\mathbf{H}$ is parallel to the $Z$ axis. 
%For uniaxial anisotropy, for example, $\mathbf{H}$ is approximated as $H_{\rm K}m_{Z}\mathbf{e}_{Z}$, 
%where $H_{\rm K}$ is a positive quantity because the $Z$ axis is assumed to be an easy axis. 
Note that the direction of the oscillation viewed from the positive $x$ direction is counterclockwise, $\dot{\varphi}_{k}>0$ (clockwise, $\dot{\varphi}_{k}<0$), 
when the magnetization oscillates near the $\theta_{k}\simeq 0$ ($\theta_{k}\simeq \pi$) direction. 
This means that $\dot{\varphi}_{k^{\prime}}\cos\theta^{\prime}$ in Eq. (\ref{eq:LLG_varphi_tmp}) is approximated to $+|\omega|$. 
Then, we obtain the following equation from Eq. (\ref{eq:LLG_varphi_tmp}), 
\begin{equation}
  \frac{d \Delta\varphi}{dt}
  \sim
  -\alpha^{\prime}
  |\omega|
  \sin \Delta\varphi. 
  \label{eq:phase_easy}
\end{equation}
It is known that the in-phase state, $\Delta\varphi=0$, is the stable fixed point of Eq. (\ref{eq:phase_easy}), 
whereas the antiphase state, $\Delta\varphi=\pi$, is an unstable fixed point \cite{strogatz01}. 
Therefore, the in-phase synchronization is excited for the oscillation around the easy axis, 
as shown in Figs. \ref{fig:fig2}(b) and \ref{fig:fig2}(c). 

% ===================================================================================================================================================================================== %

Next, let us consider the oscillation around the hard axis shown in Fig. \ref{fig:fig3}(a). 
Note that the direction of the oscillation in this case is clockwise (counterclockwise) 
when the magnetization oscillates near the $+\mathbf{e}_{z}$ ($-\mathbf{e}_{z}$) direction, 
contrary to the oscillation around the easy axis, where the direction of the oscillation is opposite. 
This difference originates from the negative sign in the demagnetization field. 
As a result, $\dot{\varphi}_{k^{\prime}}\cos\theta^{\prime}$ in Eq. (\ref{eq:LLG_varphi_tmp}) is approximated to $-|\omega|$. 
Therefore, the phase difference obeys 
\begin{equation}
  \frac{d \Delta\varphi}{dt}
  \sim
  \alpha^{\prime}
  |\omega|
  \sin \Delta\varphi. 
  \label{eq:phase_hard}
\end{equation}
The stable fixed point of Eq. (\ref{eq:phase_hard}) is the antiphase state, 
whereas the in-phase state corresponds to an unstable fixed point. 
Therefore, the antiphase synchronization is excited for the oscillation around the hard axis in the small amplitude (large current) limit, 
as shown in Fig. \ref{fig:fig3}(c). 
%We remind the readers that the small amplitude limit in this case corresponds to a large current limit. 

% ===================================================================================================================================================================================== %

Regarding these discussions, 
the reason why the phase difference between STOs becomes in-phase for the oscillation around the easy axis, 
whereas it becomes antiphase for the oscillation around the hard axis in the small amplitude limit, 
is related to the difference of the oscillation directions around the easy and hard axes. 
%The change of the stable phase difference to the in-phase found in the large amplitude (small current) limit of the oscillation around the hard axis 
%might be related to the nonlinearity of the STOs \cite{slavin09}. 

% ===================================================================================================================================================================================== %

% ===================================================================================================================================================================================== %

% ===================================================================================================================================================================================== %

\begin{figure}%[p]
\centerline{\includegraphics[width=1.0\columnwidth]{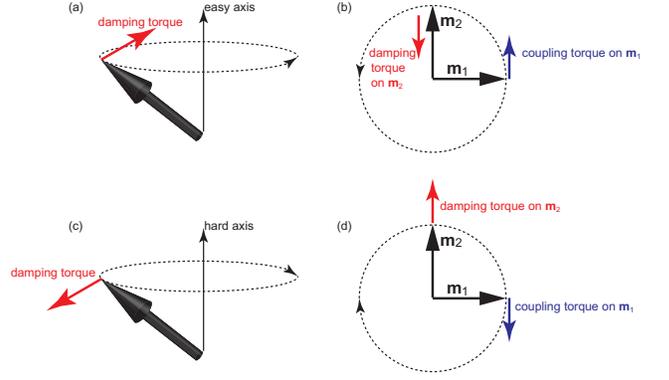}}%\vspace{-3.0ex}
\caption{
        (a) Schematic view of the magnetization oscillation around the easy axis. 
            The black arrow is the magnetization, whereas the red arrow indicates the direction of the damping torque. 
            The dotted arrow indicates the direction of the oscillation. 
        (b) The top view of the oscillations of two magnetizations around the easy axis. 
            The directions of the damping torque on $\mathbf{m}_{2}$ and the coupling torque on $\mathbf{m}_{1}$ are shown by the red and blue arrows, respectively. 
        (c) Schematic view of the magnetization oscillation around the hard axis. 
        (d) The top view of the oscillations of two magnetizations around the hard axis. 
%         \vspace{-3ex}
}
\label{fig:fig7}
\end{figure}

% ===================================================================================================================================================================================== %

% ===================================================================================================================================================================================== %

\subsection{Another approach to phase relation from schematic picture}

The above conclusions may be understood in a different manner. 
Figure \ref{fig:fig7}(a) is a schematic view of the magnetization precession around the easy axis. 
As mentioned above, 
the direction of the oscillation viewed from the positive $x$ direction is counterclockwise (clockwise) 
when the magnetization oscillates near the $+\mathbf{e}_{x}$ ($-\mathbf{e}_{x}$) direction. 
For both cases, the relative direction between the easy axis, the magnetization, and the oscillation direction is given by Fig. \ref{fig:fig7}(a). 
Figure \ref{fig:fig7}(a) also shows the direction of the damping torque, which points to the direction of the easy axis. 
Figure \ref{fig:fig7}(b) shows a top view of the oscillation trajectories of two magnetizations. 
Since the coupling torque acting on $\mathbf{m}_{k}$ due to the spin pumping points to the opposite direction to the damping torque acting on $\mathbf{m}_{k^{\prime}}$ ($k^{\prime}\neq k$), 
as described in Eq. (\ref{eq:tau_SP_tserkovnyak}), 
the direction of the coupling torque acting on $\mathbf{m}_{1}$ is given by the blue arrow in Fig. \ref{fig:fig7}(b). 
We note that the direction of the coupling torque forces $\mathbf{m}_{1}$ to move to the parallel direction of $\mathbf{m}_{2}$. 
A similar discussion holds for the coupling torque acting on $\mathbf{m}_{2}$. 
As a result, the coupling torques lead to the in-phase synchronization. 

% ===================================================================================================================================================================================== %

On the other hand, when the magnetization oscillates around the hard axis, 
the relative direction between the hard axis, the magnetization, and the oscillation direction is given by Fig. \ref{fig:fig7}(c). 
Note that the damping torque points to the opposite direction to the hard axis. 
The direction of the coupling torque in this case is schematically shown in Fig. \ref{fig:fig7}(d). 
As mentioned earlier, the coupling torque acting on $\mathbf{m}_{1}$ points to the opposite direction to the damping torque of $\mathbf{m}_{2}$. 
Then, the coupling torque forces the magnetization to move to the opposite direction of $\mathbf{m}_{2}$, as can be seen from Fig. \ref{fig:fig7}(d). 
Therefore, the coupling torques lead to the antiphase synchronization. 

% ===================================================================================================================================================================================== %

% ===================================================================================================================================================================================== %

% ===================================================================================================================================================================================== %

\section{Summary}
\label{sec:Summary}

In conclusion, a theoretical formalism was developed to self-consistently solve the LLG equation in two ferromagnets 
and the spin transport equation originated from spin pumping. 
Based on the formalism, the coupled magnetization dynamics in the STOs were studied 
for the oscillations around the easy and hard axes. 
It was found that the spin pumping leads to an in-phase synchronization of the magnetizations for the oscillation around the easy axis. 
For an out-of-plane self-oscillation, on the other hand, 
the spin pumping results in an in-phase synchronization for a small current region, 
whereas an antiphase synchronization is excited for a large current region. 
An analytical theory based on the phase equation indicated that 
the phase difference in a steady state is related to the oscillation direction, clockwise or counterclockwise, of the magnetizations. 

% ===================================================================================================================================================================================== %

\section*{Acknowledgement}

The author is grateful to Sumito Tsunegi, Takehiko Yorozu, and Hitoshi Kubota for valuable discussions. 
The author is also thankful to Satoshi Iba, Aurelie Spiesser, Hiroki Maehara, and Ai Emura 
for their support and encouragement. 
This work was supported by JSPS KAKENHI Grant-in-Aid for Young Scientists (B) 16K17486.

% ===================================================================================================================================================================================== %

\appendix

% ===================================================================================================================================================================================== %

% ===================================================================================================================================================================================== %

\section{Definition of the longitudinal conductance $g_{k}^{*}$}
\label{sec:AppendixA}

In this section, we show the examples of the longitudinal conductance $g_{k}^{*}$ in the main text. 
We note that the spin current at the F${}_{k}$/N interface due to the spin accumulation is given by \cite{brataas01} 
\begin{equation}
\begin{split}
  \mathbf{I}_{\rm s}^{{\rm F}_{k}\to{\rm N}}
  =&
  \frac{1}{4\pi}
  \left[
    \frac{(1-p_{g}^{2})g_{k}}{2}
    \mathbf{m}_{k}
    \cdot
    \left(
      \bm{\mu}_{{\rm F}_{k}}
      -
      \bm{\mu}_{\rm N}
    \right)
    \mathbf{m}_{k}
  \right.
\\
  &
  \left.
    -
    g_{{\rm r}(k)}
    \mathbf{m}_{k}
    \times
    \left(
      \bm{\mu}_{\rm N}
      \times
      \mathbf{m}_{k}
    \right)
  \right].
  \label{eq:spin_current_FN_orig}
\end{split}
\end{equation}
The interface conductance, $g_{k}=g^{\uparrow}+g^{\downarrow}$ is the sum of the conductances of the spin-up and spin-down electrons 
and is related to the interface resistance $r_{k}$ via $g_{k}=(h/e^{2})S/r_{k}$. 
The spin polarization of the conductance $g_{k}$ is denoted as $p_{g}=(g^{\uparrow}-g^{\downarrow})/(g^{\uparrow}+g^{\downarrow})$. 
The spin accumulation in the F${}_{k}$ and N layers are denoted as $\bm{\mu}_{{\rm F}_{k}}$ and $\bm{\mu}_{\rm N}$, respectively. 
For simplicity, we assume that the penetration depth of the transverse spin current in the ferromagnet is sufficiently short compared with the thickness of the ferromagnet, 
and therefore, the spin accumulation in the ferromagnet is parallel to the magnetization \cite{slonczewski96,stiles02,zhang04,taniguchi08}. 

% ===================================================================================================================================================================================== %

When the ferromagnet is an insulator, i.e., $r_{k} \to \infty$, the spin accumulation is not generated inside the ferromagnet. 
In this case, $g_{k}^{*} \to 0$. 
When the ferromagnet is a metal, on the other hand, $g_{k}^{*}$ includes the terms related to the diffusion of the longitudinal spin accumulation inside the ferromagnet. 
For example, in the ferromagnetic/nonmagnetic/ferromagnetic trilayer, 
the spin accumulation in the ferromagnet F${}_{k}$ is given by 
\begin{equation}
  \bm{\mu}_{{\rm F}_{k}}
  =
  -\frac{4\pi}{g_{{\rm sd}(k)} \sinh(d/\lambda_{\rm sd})}
  \left(
    \mathbf{m}_{k}
    \cdot
    \mathbf{I}_{\rm s}^{\rm total}
  \right)
  \cosh
  \left(
    \frac{z}{\lambda_{\rm sd}}
  \right)
  \mathbf{m}_{k},
  \label{eq:spin_accumulation_F_example}
\end{equation}
where $d$ and $\lambda_{\rm sd}$ are the thickness and the spin diffusion length of the ferromagnet, respectively. 
Here, we introduce 
\begin{equation}
  \frac{g_{{\rm sd}(k)}}{S}
  =
  \frac{h(1-\beta^{2})}{2e^{2}\rho \lambda_{\rm sd}},
\end{equation}
where $\rho$ is the resistivity of the ferromagnet, and $\beta$ is its spin polarization. 
In Eq. (\ref{eq:spin_accumulation_F_example}), we assume that the ferromagnet F${}_{k}$ lies in the region of $0 \le z \le d$, 
and is connected to the nonmagnet at $z=d$. 
The spin current at $z=0$ is zero, whereas $\mathbf{I}_{\rm s}^{\rm total}=\mathbf{I}_{\rm s}^{{\rm pump}(k)}+\mathbf{I}_{\rm s}^{{\rm F}_{k} \to {\rm N}}$ is 
the total spin current at the F${}_{k}$/N interface ($z=d$). 
The fact that a term $(\mathbf{m}_{k}\cdot\mathbf{I}_{\rm s}^{\rm total})\mathbf{m}_{k}$ appears in Eq. (\ref{eq:spin_accumulation_F_example}) is 
due to the assumption that the transverse component of the spin current is absorbed by the ferromagnet at the interface. 
Using Eqs. (\ref{eq:spin_current_FN_orig}) and (\ref{eq:spin_accumulation_F_example}), 
$g_{k}^{*}$ in the trilayer is given by \cite{tserkovnyak03} 
\begin{equation}
  \frac{1}{g_{k}^{*}}
  =
  \frac{2}{(1-p_{g}^{2})g_{k}}
  +
  \frac{1}{g_{{\rm sd}(k)} \tanh(d/\lambda_{\rm sd})}.
  \label{eq:g_star}
\end{equation}
We note that the spin accumulations in the ferromagnet and nonmagnet at the F${}_{k}$/N interface are discontinuous, 
i.e., $\bm{\mu}_{{\rm F}_{k}}-\bm{\mu}_{\rm N}$ is nonzero. 
In the present model, where the spin accumulation in the nonmagnet is assumed to be spatially uniform, 
the solution of $\bm{\mu}_{\rm N}$ is given by Eq. (\ref{eq:accumulation_sol}). 
On the other hand, the solution of $\bm{\mu}_{{\rm F}_{k}}$ is given by Eq. (\ref{eq:spin_accumulation_F_example}), 
where the total spin current at the F${}_{k}$/N interface is obtained from Eqs. (\ref{eq:spin_current_FN_orig}) and (\ref{eq:accumulation_sol}). 
The discontinuity of the spin accumulations $\bm{\mu}_{{\rm F}_{k}}$ and $\bm{\mu}_{\rm N}$ originates from 
the spin pumping, interface resistance, and absorption of the transverse spin current by the ferromagnet. 

% ===================================================================================================================================================================================== %

As can be seen in the above discussions, the explicit form of $g_{k}^{*}$ depends on the boundary condition of the spin accumulation. 
When two ferromagnets are replaced by STOs, the explicit form of $g_{k}^{*}$ will be complex 
because additional nonmagnets and ferromagnets are attached to the oscillating layers. 
The role of such additional layers is, however, only the renormalization of $g_{k}^{*}$ and $g_{{\rm r}(k)}$. 
The conventional STO usually consists of a metallic free layer. 
In addition, when the STO consists of a magnetic tunnel junction (MTJ), 
the trilayer model consisting of the oscillating free layers and nonmagnetic connector 
is expected to work well due to the presence of the tunneling barriers. 
Therefore, in the main text, we use Eq. (\ref{eq:g_star}) as a definition of $g_{k}^{*}$, for simplicity. 

% ===================================================================================================================================================================================== %

% ===================================================================================================================================================================================== %

\section{The explicit form of the matrix $\mathsf{L}$}
\label{sec:AppendixB}

In this section, we show the explicit forms of the components of the matrix $\mathsf{L}$. 

% ===================================================================================================================================================================================== %

\subsection{Derivation of the matrix $\mathsf{L}$}

%As written in the main text, we assume that the distance between the ferromagnets is sufficiently shorter than the width of the nonmagnet, 
%and therefore, the spin current is conserved inside the nonmagnet. 
Equation (\ref{eq:conservation}) in the Cartesian coordinate can be rewritten as 
\begin{equation}
  \mathsf{K}
  \bm{\mu}_{\rm N}
  =
  \hbar 
  \sum_{k=1,2}
  g_{{\rm r}(k)}
  \mathbf{m}_{k}
  \times
  \dot{\mathbf{m}}_{k},
\end{equation}
where the $3\times 3$ matrix $\mathsf{K}$ is given by 
\begin{widetext}
\begin{equation}
  \mathsf{K}
  =
  \sum_{k=1,2}
  \begin{pmatrix}
    g_{{\rm r}(k)}-(g_{{\rm r}(k)}-g_{k}^{*})m_{kx}^{2} &
    -(g_{{\rm r}(k)}-g_{k}^{*})m_{kx}m_{ky} & 
    -(g_{{\rm r}(k)}-g_{k}^{*})m_{kx}m_{kz} \\
    -(g_{{\rm r}(k)}-g_{k}^{*})m_{ky}m_{kx} &
    g_{{\rm r}(k)}-(g_{{\rm r}(k)}-g_{k}^{*})m_{ky}^{2} &
    -(g_{{\rm r}(k)}-g_{k}^{*})m_{ky}m_{kz} \\
    -(g_{{\rm r}(k)}-g_{k}^{*})m_{kz}m_{kx} &
    -(g_{{\rm r}(k)}-g_{k}^{*})m_{kz}m_{ky} &
    g_{{\rm r}(k)}-(g_{{\rm r}(k)}-g_{k}^{*})m_{kz}^{2}
  \end{pmatrix}.
\end{equation}
\end{widetext}
Therefore, the spin accumulation $\bm{\mu}_{\rm N}$ in the nonmagnet is obtained as 
\begin{widetext}
\begin{equation}
\begin{split}
  \bm{\mu}_{\rm N}
  &=
  \hbar 
  \mathsf{K}^{-1}
  \sum_{k=1,2}
  g_{{\rm r}(k)}
  \mathbf{m}_{k}
  \times
  \dot{\mathbf{m}}_{k}
\\
  &=
  \hbar 
  \sum_{k=1,2}
  g_{{\rm r}(k)}
  \begin{pmatrix}
    (K_{12}^{-1}m_{kz}-K_{13}^{-1}m_{ky}) \dot{m}_{kx} + (K_{13}^{-1}m_{kx}-K_{11}^{-1}m_{kz}) \dot{m}_{ky} + (K_{11}^{-1}m_{ky}-K_{12}^{-1}m_{kx}) \dot{m}_{kz} \\
    (K_{22}^{-1}m_{kz}-K_{23}^{-1}m_{ky}) \dot{m}_{kx} + (K_{23}^{-1}m_{kx}-K_{21}^{-1}m_{kz}) \dot{m}_{ky} + (K_{21}^{-1}m_{ky}-K_{22}^{-1}m_{kx}) \dot{m}_{kz} \\
    (K_{32}^{-1}m_{kz}-K_{33}^{-1}m_{ky}) \dot{m}_{kx} + (K_{33}^{-1}m_{kx}-K_{31}^{-1}m_{kz}) \dot{m}_{ky} + (K_{31}^{-1}m_{ky}-K_{32}^{-1}m_{kx}) \dot{m}_{kz}
  \end{pmatrix}, 
  \label{eq:accumulation_sol_tmp}
\end{split}
\end{equation}
\end{widetext}
where $K^{-1}_{ab}$ ($a,b=1$,$2$,$3$ or $x$,$y$,$z$) is a component of a $3\times 3$ matrix $\mathsf{K}^{-1}$, 
which is the inverse matrix of $\mathsf{K}$. 
When two ferromagnets have an identical property, i.e., $g_{1}^{*}=g_{2}^{*}$ and $g_{\rm r(1)}=g_{\rm r(2)}$, 
the explicit form of $K_{ab}^{-1}$ is given by Eq. (\ref{eq:matrix_K_inv_component}). 
%\begin{equation}
%\begin{split}
%  K_{ab}^{-1}
%  &=
%  \frac{4 g_{\rm r}g^{*} \delta_{ab} + 2 g_{\rm r}(g_{\rm r}-g^{*})(m_{1a}m_{1b}+m_{2a}m_{2b})+(g_{\rm r}-g^{*})^{2}\mathbf{e}_{a}\cdot(\mathbf{m}_{1}\times\mathbf{m}_{2})\mathbf{e}_{b}\cdot(\mathbf{m}_{1}\times\mathbf{m}_{2})}
%    {2g_{\rm r}[(g_{\rm r}+g^{*})^{2} - (g_{\rm r}-g^{*})^{2}(\mathbf{m}_{1}\cdot\mathbf{m}_{2})^{2}]},
%\\
%  &=
%  \frac{(1-\nu^{2}) \delta_{ab} + (1+\nu)\nu(m_{1a}m_{1b}+m_{2a}m_{2b})+\nu^{2}\mathbf{e}_{a}\cdot(\mathbf{m}_{1}\times\mathbf{m}_{2})\mathbf{e}_{b}\cdot(\mathbf{m}_{1}\times\mathbf{m}_{2})}  
%    {2g_{\rm r}[1 - \nu^{2}(\mathbf{m}_{1}\cdot\mathbf{m}_{2})^{2}]},
%\end{split}
%\end{equation}
%where $\nu$ is defined as \cite{tserkovnyak03}
%\begin{equation}
%  \nu
%  =
%  \frac{g_{\rm r}-g^{*}}{g_{\rm r}+g^{*}}. 
%\end{equation}

Equation (\ref{eq:accumulation_sol_tmp}) can be generally rewritten as 
\begin{equation}
  \bm{\mu}_{\rm N}
  =
  \hbar
  \sum_{\ell=1,2}
  \mathsf{N}_{(\ell)}
  \dot{\mathbf{m}}_{\ell}
  \label{eq:accumulation_sol}
\end{equation}
where $\mathsf{N}_{(\ell)}$ is a $3\times 3$ matrix given by 
\begin{equation}
%  \mathsf{N}_{(\ell)}
  N_{(\ell)ab}
  =
  g_{{\rm r}(\ell)}
  \epsilon_{bcd}
  K_{ac}^{-1}
  m_{\ell d}.
%  =
%  g_{{\rm r}(\ell)}
%  \begin{pmatrix}
%    K_{12}^{-1}m_{\ell z}-K_{13}^{-1}m_{\ell y} &
%    K_{13}^{-1}m_{\ell x}-K_{11}^{-1}m_{\ell z} &
%    K_{11}^{-1}m_{\ell y}-K_{12}^{-1}m_{\ell x} \\
%    K_{22}^{-1}m_{\ell z}-K_{23}^{-1}m_{\ell y} &
%    K_{23}^{-1}m_{\ell x}-K_{21}^{-1}m_{\ell z} & 
%    K_{21}^{-1}m_{\ell y}-K_{22}^{-1}m_{\ell x} \\
%    K_{32}^{-1}m_{\ell z}-K_{33}^{-1}m_{\ell y} &
%    K_{33}^{-1}m_{\ell x}-K_{31}^{-1}m_{\ell z} &
%    K_{31}^{-1}m_{\ell y}-K_{32}^{-1}m_{\ell x} 
%  \end{pmatrix}. 
\end{equation}
Equation (\ref{eq:matrix_N_component}) is reproduced by assuming that the value of the mixing conductance is the same for both F${}_{1}$/N and F${}_{2}$/N interface. 
The components of the $3\times 3$ matrix $\mathsf{N}_{(k.\ell)}^{\prime}$ are given by Eq. (\ref{eq:matrix_Np_component}). 
%\begin{equation}
%  N_{(k,\ell)ab}^{\prime}
%  =
%  \sum_{c=x,y,z}
%  m_{kc}
%  m_{ka}
%  N_{(\ell)cb}. 
%\end{equation}
Using $\mathsf{N}_{(k,\ell)}^{\prime}$, we find that 
\begin{equation}
  \left(
    \mathbf{m}_{k}
    \cdot
    \bm{\mu}_{\rm N}
  \right)
  \mathbf{m}_{k}
  =
  \hbar
  \sum_{\ell=1,2}
  \mathsf{N}_{(k,\ell)}^{\prime}
  \dot{\mathbf{m}}_{\ell}.
  \label{eq:accumulation_projection}
\end{equation}

We should remind the reader that the spin torque due to the spin pumping, 
\begin{equation}
  \bm{\tau}_{k}^{\rm SP}
  =
  \frac{\gamma}{MV}
  \mathbf{m}_{k}
  \times
  \left[
    \left(
      \mathbf{I}_{\rm s}^{{\rm pump}(k)}
      +
      \mathbf{I}_{\rm s}^{{\rm F}_{k}\to{\rm N}}
    \right)
    \times
    \mathbf{m}_{k}
  \right],
  \label{eq:spin_pumping_torque_approx}
\end{equation}
consists of two terms. 
The first contribution, 
\begin{equation}
  \frac{\gamma}{MV}
  \mathbf{m}_{k}
  \times
  \left(
    \mathbf{I}_{\rm s}^{{\rm pump}(k)}
    \times
    \mathbf{m}_{k}
  \right)
  =
  \frac{\gamma\hbar g_{{\rm r}(k)}}{4\pi MV}
  \mathbf{m}_{k}
  \times
  \dot{\mathbf{m}}_{k},
\end{equation}
gives an additional damping $\alpha_{k}^{\prime}$ given by 
\begin{equation}
  \alpha_{k}^{\prime}
  =
  \frac{\gamma\hbar g_{{\rm r}(k)}}{4\pi MV}. 
\end{equation}
On the other hand, the second contribution, 
\begin{equation}
\begin{split}
  \frac{\gamma}{MV}
  \mathbf{m}_{k}
  \times
  \left(
    \mathbf{I}_{\rm s}^{{\rm F}_{k} \to {\rm N}}
    \times
    \mathbf{m}_{k}
  \right)
  &=
  -\frac{\gamma g_{{\rm r}(k)}}{4\pi MV}
  \mathbf{m}_{k}
  \times
  \left(
    \bm{\mu}_{\rm N}
    \times
    \mathbf{m}_{k}
  \right)
\\
  &=
  -\frac{\gamma g_{{\rm r}(k)}}{4\pi MV}
  \left[
    \bm{\mu}_{\rm N}
    -
    \left(
      \mathbf{m}_{k}
      \cdot
      \bm{\mu}_{\rm N}
    \right)
    \mathbf{m}_{k}
  \right],
\end{split}
\end{equation}
can be rewritten as 
\begin{equation}
  \frac{\gamma}{MV}
  \mathbf{m}_{k}
  \times
  \left(
    \mathbf{I}_{\rm s}^{{\rm F}_{k} \to {\rm N}}
    \times
    \mathbf{m}_{k}
  \right)
  =
  -\alpha_{k}^{\prime}
  \sum_{\ell=1,2}
  \left[
    \mathsf{N}_{(\ell)}
    -
    \mathsf{N}_{(k,\ell)}^{\prime}
  \right]
  \dot{\mathbf{m}}_{\ell},
\end{equation}
where we use Eq. (\ref{eq:accumulation_sol}) and (\ref{eq:accumulation_projection}). 

In summary, the LLG equation, Eq. (\ref{eq:LLG}), can be rewritten as 
\begin{equation}
  \mathsf{L}
  \begin{pmatrix}
    \dot{\mathbf{m}}_{1} \\
    \dot{\mathbf{m}}_{2}
  \end{pmatrix}
  =
  \begin{pmatrix}
    -\gamma \mathbf{m}_{1} \times \mathbf{H}_{1} + \bm{\tau}_{1}^{\rm STT} \\
    -\gamma \mathbf{m}_{2} \times \mathbf{H}_{2} + \bm{\tau}_{2}^{\rm STT}
  \end{pmatrix}, 
  \label{eq:LLG_SP_approx}
\end{equation}
where the explicit form of the $6 \times 6$ matrix $\mathsf{L}$ is given by 
\begin{equation}
\begin{split}
  \mathsf{L}
  =&
  \hat{I}
  +
  (\alpha_{0}+\alpha_{1}^{\prime})
  \begin{pmatrix}
    0 & m_{1z} & -m_{1y} & 0 & 0 & 0 \\
    -m_{1z} & 0 & m_{1x} & 0 & 0 & 0 \\
    m_{1y} & -m_{1x} & 0 & 0 & 0 & 0 \\
    0 & 0 & 0 & 0 & 0 & 0 \\
    0 & 0 & 0 & 0 & 0 & 0 \\
    0 & 0 & 0 & 0 & 0 & 0
  \end{pmatrix}
\\
  & +
  (\alpha_{0}+\alpha_{2}^{\prime})
  \begin{pmatrix}
    0 & 0 & 0 & 0 & 0 & 0 \\
    0 & 0 & 0 & 0 & 0 & 0 \\
    0 & 0 & 0 & 0 & 0 & 0 \\
    0 & 0 & 0 & 0 & m_{2z} & -m_{2y} \\
    0 & 0 & 0 & -m_{2z} & 0 & m_{2x} \\
    0 & 0 & 0 & m_{2y} & -m_{2x} & 0
  \end{pmatrix}
\\
  &+
  \begin{pmatrix}
    \alpha_{1}^{\prime}[\mathsf{N}_{(1)} - \mathsf{N}_{(1,1)}^{\prime}] & \alpha_{1}^{\prime}[\mathsf{N}_{(2)} - \mathsf{N}_{(1,2)}^{\prime}] \\
    \alpha_{2}^{\prime}[\mathsf{N}_{(1)} - \mathsf{N}_{(2,1)}^{\prime}] & \alpha_{2}^{\prime}[\mathsf{N}_{(2)} - \mathsf{N}_{(2,2)}^{\prime}]
  \end{pmatrix},
  \label{eq:matrix_L_approx}
\end{split}
\end{equation}
where $\hat{I}$ is the $6 \times 6$ unit matrix. 
The readers are to be reminded that $\mathsf{N}_{(\ell)}$ and $\mathsf{N}_{(k,\ell)}$ are the $3\times 3$ matrices. 
In the main text, we assume that the material parameters of two ferromagnets are identical 
and remove the suffix $k(=1,2)$ from $g_{k}^{*}$, $g_{{\rm r}(k)}$ and $\alpha_{k}^{\prime}$. 

% ===================================================================================================================================================================================== %

% ===================================================================================================================================================================================== %

\section{The role of angular dependence of spin torque}
\label{sec:AppendixC}

In the numerical simulation, we use Eq. (\ref{eq:STT}) as a spin torque. 
In general, however, the angular dependence of the spin torque is described by a more complex function. 
For example, in a symmetric MTJ, the spin torque is given by \cite{slonczewski05} 
\begin{equation}
  \bm{\tau}_{k}^{\rm STT}
  =
  -\frac{\gamma \hbar \eta j}{2e(1+\lambda \mathbf{m}_{k}\cdot\mathbf{p})Md}
  \mathbf{m}_{k}
  \times
  \left(
    \mathbf{p}
    \times
    \mathbf{m}_{k}
  \right),
  \label{eq:STT_Slonczewski}
\end{equation}
where $\lambda=\eta^{2}$ is a dimensionless parameter. 
Two dimensionless parameters, $\eta$ and $\lambda$, in Eq. (\ref{eq:STT_Slonczewski}) characterize 
the spin polarization of the ferromagnet and the angular dependence of the magnetoresistance \cite{slonczewski05}. 
It has been known that the term $\lambda \mathbf{m}_{k}\cdot\mathbf{p}$ affects, for example, the threshold current of the self-oscillation \cite{taniguchi16}. 
For example, $J_{\rm c}$ given by Eq. (\ref{eq:Jc}) is changed to $(1+\lambda)J_{\rm c} \simeq 22$ MA/cm${}^{2}$ when a finite $\lambda=\eta^{2}$ is taken into account. 
We confirm this change of the threshold current from the numerical simulation. 
Simultaneously, however, we confirmed that the phase difference in the synchronized state is unchanged by adding the factor $\lambda \mathbf{m}_{k}\cdot\mathbf{p}$, 
i.e., the phase difference is zero for the in-plane oscillation around the easy axis, 
whereas the phase difference is changed from in-phase to antiphase by increasing the current for the out-of-plane oscillation around the hard axis. 

% ===================================================================================================================================================================================== %

The angular dependence of the spin torque in a typical ferromagnetic nanostructure is generally described by four dimensionless parameters \cite{xiao04}, 
which correspond to $\eta$ and $\lambda$ for the free and pinned layers. 
A different kind of synchronization may appear if we change the four parameters in the wide ranges of their values. 
However, such investigation is beyond the scope of this paper. 

% ===================================================================================================================================================================================== %

% ===================================================================================================================================================================================== %

% ===================================================================================================================================================================================== %

% ===================================================================================================================================================================================== %

% ===================================================================================================================================================================================== %

% ===================================================================================================================================================================================== %

% ===================================================================================================================================================================================== %

%\bibliography{biblist}

% ================================================================================================================================================================================= %

\end{document}